\documentstyle[preprint,aps]{revtex}  
\begin{document} 
\draft  
\title{Charged right-handed currents in the leptoquark-bilepton  
flavordynamics chiral models}     
\author{A. A. Machado and F. Pisano}    
\address{Departamento de F\'\i sica, Universidade Federal do Paran\'a,    
81531-990 Curitiba, PR, Brazil}  
\date{\today}  
\maketitle 
\begin{abstract} 
Flavordynamics chiral models of leptoquark fermions and bilepton   
gauge bosons with masses up to about 1 TeV,        
although coincident  at low energy with the standard model,   
address the question of fermion mass hierarchy and explain the   
flavor question.  
The presence of charged right-handed weak currents coupled to   
bileptons, which we point out herein, is another remarkable  
feature of these chiral theories.      
\end{abstract} 
\bigskip 
\newpage    
\par   
The nature of weak interaction appears to be intimately connected   
with the fundamental questions like whether neutrinos have finite  
mass~\cite{kamioka98} and whether the right-handed weak charged 
current exists. These are the relevant issues to provide evidence   
for possible new physics beyond the standard electroweak model.   
The current-current $V-A$ theory~\cite{v-a}   
of charged-current weak interactions was motivated by the    
$\gamma_5$ chirality invariance of the Weyl equation of a   
massless neutrino and its extension to charged fermions.        
Neutrinos were assumed to be massless so that they are in the   
definite helicity state (Weyl neutrinos) to assure the $V-A$   
structure of leptonic current. In the standard model the  
fermions are introduced as left-handed doublets and    
right-handed singlets under the SU(2) weak isospin        
symmetry group.   
Right-handed constituents of matter do not interact with the  
charged $W$ gauge bosons and thus there are no right-handed  
charged currents. The $V-A$ Lorentz structure guarantees maximal  
parity violation. Only left-handed chiral fermions are   
coupled in the weak interactions governed by the charged  
currents. In the standard model scheme   
the photonic and gluonic vertices are vectorlike ($V$), the  
$W^\pm$ vertices are purely $V-A$, whereas the $Z^0$ vertices   
involve both $V-A$ and $V+A$ structures.     
\par  
Experimental data are consistent with the standard  
model~\cite{rosner98}   
so that some theoretical extensions are motivated by attempting   
to understand features that are inserted into the standard   
flavordynamics but not explained by it. Considering the lightest   
known particles as the sector in which a symmetry is manifested,   
it is interesting that the lepton sector could be the part of the   
model determining new approximate symmetries.    
There is an electroweak   
model~\cite{adler89,pp92,framp92,mpp93,pleton93,ng94}    
in which there appear SU(2)$_L$-doublet gauge bosons      
$(V^\mp, U^{\mp\mp})$ carrying lepton number $L=\pm 2$. We refer   
to these gauge bosons as bileptons~\cite{cuypers}. The gauge   
group of the standard model is extended to   
$G_0\equiv$ SU(3)$_C$ $\otimes$ $G_W$ $\otimes$ U(1)$_N$ with     
$G_W=$ SU(3)$_L$ but if each   
family of fermions is treated separately and lepton  
electric charges are only $0$ and $\pm 1$ then $G_W=$ SU(4)$_L$   
is the highest weak isospin symmetry which maintain the    
bilepton character~\cite{su4}. Such   
bileptons appeared first in stable-proton grand unified   
theories~\cite{framplee} with non-chiral fermions and         
anomaly cancellation through mirror fermions. However, in the  
$G_0$ bilepton models, the chiral anomalies between families of  
fermions cancel with a matching of the number of families and  
the number of color degrees of freedom, solving the flavor  
question~\cite{framp92,eu96}. Furthermore, the family  
horizontal symmetry, under which the families transform by a     
non-trivial representation, can explain the fermion  
mass hierarchy~\cite{fram95}.     
\par   
Different representation contents are determined by  
embedding the electric charge operator 
\begin{equation}  
\frac{{\cal Q}}{e} = \frac{1}{2} 
(\lambda^L_3 + \xi\lambda^L_8 + \zeta\lambda^L_{15}) + N              
\label{um}
\end{equation}
in the neutral generators of $G_W$$\otimes$U(1)$_N$.   
The parameters $\xi$ and $\zeta$ distinguish different embeddings    
and $N$ is the new U(1) charge.        
Let us consider the $G_W=$ SU(3)$_L$ chiral bilepton model with  
the embedding parameters $\xi=-\sqrt 3$, $\zeta = 0$~\cite{pp92}.   
All charged lepton degrees of freedom, $l_L$, $(l_R)^c$;   
$l=e,\mu,\tau$, where $l^c\equiv C\bar l^T$ is the charge conjugate  
field corresponding to $l$, belong to the same flavor triplet     
\begin{equation} 
f_{lL} = (\nu_l,\,l,\,l^c)_L \sim ({\bf 1},{\bf 3},0).    
\label{dois}
\end{equation}    
Hence there are no right-handed flavor singlets. If we admit that   
right-handed neutrinos do exist, it is possible to build a model  
in which $\nu^c_l$, $\nu_l$ and $l$ are in the same multiplet of  
SU(3)$_L$~\cite{mpp93,long96}. However, if right-handed neutrinos  
are introduced it is an interesting possibility to have  
$\nu_l$, $l$, $\nu^c_l$, and $l^c$ in the same multiplet 
\begin{equation}    
f_{lL} = (\nu_l,\,l,\,\nu^c_l,\,l^c)_L \sim ({\bf 1},{\bf 4},0)    
\label{tres}
\end{equation}
of $G_W=$ SU(4)$_L$ bilepton theory~\cite{su4}.         
In the SU(3)$_L$ model the quark multiplets transform as           
$Q_{1L} = (u_1,\,d_1,\,J_1)_L \sim ({\bf 3},{\bf 3},2/3)$,     
$Q_{2L} = (J_2,\,u_2,\,d_2)_L \sim ({\bf 3},{\bf 3}^*,-1/3)$,    
$Q_{3L} = (J_3,\,u_3,\,d_3)_L \sim ({\bf 3},{\bf 3}^*,-1/3)$,     
and the right-handed fields are introduced as SU(3)$_L$ singlets,   
$u_{iR}\sim ({\bf 3},{\bf 1},2/3)$,     
$d_{iR}\sim ({\bf 3},{\bf 1},-1/3)$,       
$J_{1R}\sim ({\bf 3},{\bf 1},5/3)$,       
$J_{2,3R}\sim ({\bf 3},{\bf 1},-4/3)$. 
In order to complete the quark triplets, the inclusion of exotic  
leptoquark fermions of charge $\pm 5/3$ and $\mp 4/3$ has been   
necessary. They are  
color-triplet particles which possess baryon number and lepton  
number. The $L$ and $B$ quantum numbers carried by the leptoquarks  
are $L_{J_{2,3}}=-L_{J_1}=2 $, and $B_{J_{1,2,3}}=1/3$. In the   
bilepton models with the full $G_0$ gauge symmetry two of the three  
quark families transform identically and one family    
transforms in a different representation under  
the electroweak group factor of $G_0$.   
This type of construction is anomaly-free only if  
there equal number of ${\bf 3}$ and ${\bf 3}^*$    
or ${\bf 4}$ and ${\bf 4}^*$, considering the color degrees of   
freedom, $N_C$. The number of families, $N_f$, must be divisible  
by $N_C$. Hence the simplest alternative is $N_C=N_f=3$. This novel  
method of anomaly cancellation requires that at least one family  
transforms differently from the others, thus breaking generation  
universality. Using experimental input on neutral gauge bosons  
mixing, the third quark family must be the one that is singled   
out~\cite{liu}.      
\par   
The interactions among the gauge bosons and fermions are read off  
from  
\begin{equation}  
{\cal L}_F = \overline Ri\gamma^\mu (\partial_\mu + ig_1B_\mu N_R)R + 
\overline Li\gamma^\mu (\partial_\mu + ig_1B_\mu N_L + i\frac{g_2}{2} 
\lambda^aW^a_\mu)L,  
\label{qua}  
\end{equation}  
where $R$ represents any right-handed isosinglet and $L$ any  
left-handed triplet or ${\bf 4}$-plet and $g_{1,2}$ are the gauge  
couplings associated to U(1)$_N$ and $G_W$, respectively.   
For leptons the charged-current interactions in terms of      
symmetry eigenstates are  
\begin{equation}  
{\cal L}_l^{\rm CC}[{\rm SU(3)}_L] = -\frac{g_2}{\sqrt 2} \sum_l 
(\bar\nu_{lL}\gamma^\mu l_L W^+_\mu 
+\overline{l^c_L}\gamma^\mu\nu_{lL} V^+_\mu 
+\overline{l^c_L}\gamma^\mu l_L U^{++}_\mu) + {\rm H.c.}     
\label{cinque}
\end{equation}
within the $G_W=$ SU(3)$_L$ model and    
\begin{eqnarray}  
{\cal L}_l^{\rm CC}[{\rm SU(4)}_L] = &-&\frac{g_2}{\sqrt 2} \sum_l 
(\bar\nu_{lL}\gamma^\mu l_L W^+_\mu   
+\overline{\nu_{lL}^c}\gamma^\mu l_LV^+_{1\mu} \nonumber \\ 
&+& \overline{l^c_L}\gamma^\mu\nu_{lL} V^+_{2\mu} 
+ \overline{l^c_L}\gamma^\mu\nu^c_{lL} V^+_{3\mu} + 
\overline{l^c_L}\gamma^\mu l_L U^{++}_\mu) + {\rm H.c.}   
\label{sei}
\end{eqnarray}
in the $G_W=$ SU(4)$_L$ model. The left-handed currents coupled   
to the bilepton gauge bosons $V$, $U$, and $V_2$, $V_3$ contain  
the $l^c$ charge conjugate fields. Notwithstanding these currents  
are purely right-handed for the $l$ negative charged fields,    
\begin{equation}  
\overline{l^c_L}\gamma^\mu\nu_{lL} = 
-\overline{\nu^c_{lR}}\gamma^\mu l_R,   
\label{sete}   
\end{equation}  
\begin{equation}
\overline{l^c_L}\gamma^\mu\nu^c_{lL} =   
-\bar\nu_{lR}\gamma^\mu l_R  
\label{otto}
\end{equation}
and also     
\begin{equation} 
\overline{l^c_L}\gamma^\mu l_L =   
-\overline{l^c_R}\gamma^\mu l_R,          
\label{nove}
\end{equation}
as can be cheked, for any spinor $\psi$, through the charge conjugation   
definition,            
$\psi^c=C\bar\psi^T$, $C$ being the charge conjugation matrix,     
the Dirac conjugate $\bar\psi^c=-\psi^TC^{-1}$,     
the property of Dirac matrices, $C^{-1}\gamma_\mu C= -\gamma^T_\mu$,    
and the equal-time anticommutation of fermion fields.      
We take the bar after the $C$ operation in $\bar\psi^c$.               
Up to now, right-handed currents have always been considered   
in the context of the left-right symmetric models~\cite{leftright}.   
However, it is possible to have these currents coupled to the  
bilepton gauge bosons in the chiral left-handed models.      
An explicit lepton number breakdown was proposed in        
Ref.~\cite{valle} but in that reference the lepton number-violating 
currents are coupled to the standard gauge bosons and are  
proportional to a small parameter appearing in the model.   
\par  
Lastly, note that although in the left-right symmetric theories     
the existence of $V+A$ currents ought to be related with a  
nonvanishing neutrino mass, the smallness of the neutrino mass  
and the supression of $V+A$ currents should be   
connected~\cite{senjanovic81}, in the leptoquark-bilepton    
models could be there an additional connection constraining the   
electromagnetic gauge invariance and the nature of neutrino  
mass terms~\cite{ozer}. However, as was shown recently~\cite{pt98},  
it is possible that all the vacuum expectation values of the scalar   
sextets be non-zero giving mass to the charged and neutral   
leptons while keeping the photon massless.                 

\end{document}